\documentclass[twocolumn]{jpsj3}
\usepackage{bm,amssymb,amsmath}
\usepackage{times}
\usepackage{color}
\newcommand{\simg}{\stackrel{>}{_\sim}}

\title{Hole-$s_{\pm}$ State Induced by Coexisting Ferro- and Aniferromagnetic and Aniferro-orbital Fluctuations in the Iron Pnictides}

\author{Jun Ishizuka$^1$, Takemi Yamada$^1$, Yuki Yanagi$^2$, Yoshiaki \=Ono$^1$ 
}
\inst{1. Department of Physics, Niigata University, Ikarashi, Niigata 950-2181, Japan \\
2. Department of Physics, Faculty of Science and Technology, Tokyo University of Science, Noda 278-8510, Japan 
}
\abst{
The multi-orbital Hubbard model is investigated in order to clarify the electron correlation effects and the superconductivity in the iron-based superconductor. The renormalization effects on the self-energy and the two-particle irreducible vertex function are calculated on the basis of the dynamical mean field theory. We find that the vertex function exhibits a strong renormalization with the significant orbital dependence as compared with the renormalization factor, when the antiferromagnetic and the antiferro-orbital fluctuations are comparably enhanced due to the electron and the hole Fermi surfaces nesting effects. The orbital-dependent vertex function together with the ${\bm q}\sim\bm0$ nesting between the two-hole Fermi surfaces results in the inter-orbital ferromagnetic fluctuation gradually enhanced which is expected to be observed in LiFeAs. We show that the hole-$s_{\pm}$-wave superconductivity with the sign change of the two-hole Fermi surfaces is realized by the enhanced ferromagnetic fluctuation accompanied by the antiferromagnetic fluctuation and the antiferro-orbital fluctuation.
}

\kword{dynamical mean field theory, iron-based superconductor, magnetic fluctuation, orbital fluctuation, 5-orbital Hubbard model}
\begin{document}
\maketitle

\section{Introduction}
The discovery of high-temperature superconductivity in iron-based compounds\cite{kamihara} has attracted much attention to investigate their electronic state and superconducting mechanism. The most of the 1111 and the 122 systems show the tetragonal-orthorhombic structural transition and the stripe-type antiferromagnetic (AFM) transition. Corresponding to the structural and the AFM transitions, two distinct $s$-wave pairings: the $s_{\pm}$-wave state with sign change of the gap function $\Delta({\bm k})$ between the hole and the electron Fermi surfaces (FSs) mediated by the antiferromagnetic fluctuation\cite{mazin,kuroki} and the $s_{++}$-wave state without the sign change mediated by the ferro-orbital (FO) fluctuation, which is responsible for the softening of $C_{66}$ through the mode-coupling correction\cite{onari} or the electron-phonon coupling\cite{yanagi_2}, and/or the antiferro-orbital (AFO) fluctuation\cite{yanagi_3,kontani} were proposed.

Also, spin excitation, superconducting gap structure and phase diagram varies among iron-based families. For example, unlike 1111 and 122 systems, LiFeAs is a superconductor in its stoichiometric form; any chemical substitution on the Fe site causes a reduction in the transition temperature\cite{pitcher_2}, and ordered magnetic phase or structural transition (or softening of $C_{66}$) has yet been observed\cite{tapp}. However, incommensurate-AFM fluctuations\cite{knolle} are exist, and FM fluctuation has also been observed by $\mu$SR experiment\cite{wright_2} albeit in high temperature. Its fluctuation seems to be intriguing because the first-principles band calculation for LiFeAs leads to give antiferromagnetic groundstate with orthorhombic distortion similar to the 1111 and 122 systems\cite{li_2}.

From theoretical study of superconducting gap structure, the possibility of the spin-triplet $p$-wave pairing due to ferromagnetic (FM) fluctuation in LiFeAs has been discussed in the three-orbital Hubbard model within random phase approximation (RPA) because of a bad nesting between the hole and the electron FSs\cite{brydon}. On the other hands, the hole-$s_{\pm}$-wave pairing with the sign change between the two-hole FSs, and without the sign change between the two electron FS mediated by coexistence of the AFM and the AFO fluctuations, is discussed in the realistic five-orbital Hubbard model\cite{saito} by RPA or mode-coupling theory. In addition, the orbital antiphase $s_{\pm}$-wave pairing\cite{yin_2}, with the sign change between the both two-hole and two-electron FS has been suggested by the combination of the density functional theory and the dynamical mean field theory (DMFT). It is noted that the orbital antiphase $s_{\pm}$-wave state makes nodal picture on electronlike FS in the unfolding Brillouin zone.

As the electron correlation effects together with the details of band structure are crucial for the metallic magnetisms, we investigate the realistic five-orbital Hubbard model derived from the first-principles band calculation\cite{mazin,kuroki}, by using the DMFT which enables us to take into account of the local correlation effects sufficiently. The DMFT has become almost standard for treating electronic-correlated systems in the last decade and has been able to explain some of the spectral\cite{haule,craco} and magnetic properties\cite{haule_2,hansmann,aichhorn} of iron-based superconductors. Yin and Kotliar\cite{yin_2} have proved that the dynamical magnetic susceptibility for all momenta and frequency, which requires the determination of the local irreducible vertices at DMFT level, reproduces the result of inelastic neutron scattering experiment in detail. Despite the numerous efforts, the pairing state together with the mechanism of the superconductivity is still controversial.

In our previous research\cite{ishizuka,ishizuka_2}, we pointed out that the local correlation effects affect on the possible pairing states: the magnetic-fluctuation-mediated $s_{\pm}$-wave state and the orbital-fluctuation-mediated $s_{++}$-wave state, by using the DMFT combined with the Eliashberg equation beyond the Hartree-Fock (HF) approximation and the RPA. We have found that the $s_{++}$-wave state is largely expanded relative to the RPA result, while the $s_\pm$-wave state is reduced, because the local component of the repulsive (attractive) pairing interaction are responsible for the suppression (enhancement) of the $s_\pm$($s_{++}$)-pairing. This is because the self-energy correction and vertex correction is considered in the DMFT. As is well known, the self-energy correction have been understood as describing the non Fermi-liquid behavior\cite{liebsch_2}, the orbital selective change in the band dispersion\cite{lee_3} and the orbital selective Mott transition\cite{medici_1} as observed in recent experiments\cite{ding_2,maletz,yi_2}. However, the effect of the two particle vertex correction have not been fully understood in the multi-orbital system. The purpose of this paper is to clarify the relation between the vertex correction and superconductivity in the wide parameter region.

\section{Model and Formulation}
\label{sect_model}
\begin{figure}
\begin{center}
\includegraphics[width=74mm]{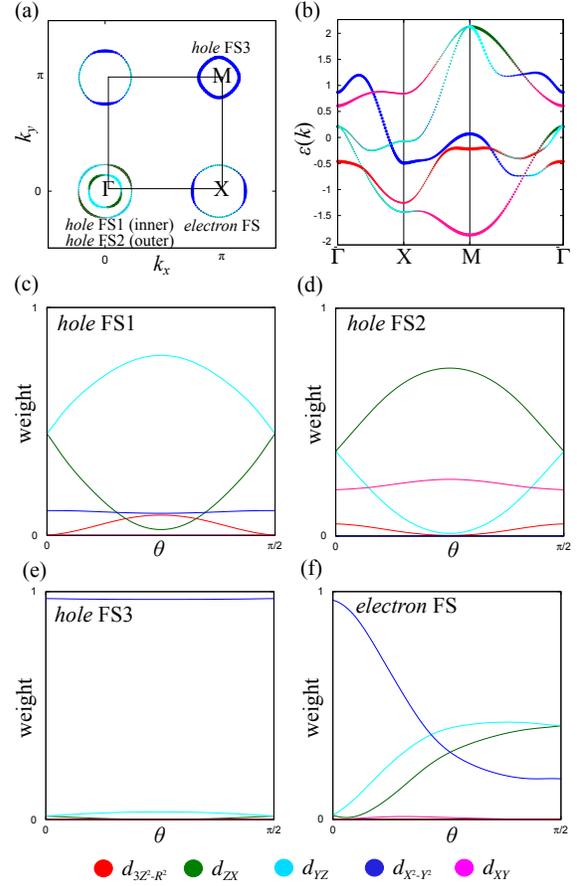}
\caption{(Color online)
(a) The FSs of the five-orbital model (b) The dispersion of the band structure. (c)-(f) The weight of the each $d$ orbitals on the FSs. The horizontal axis is $\theta=\tan^{-1}(k_y/k_x)$. We number the orbitals as follows: (1) $d_{3Z^2-R^2}$ (red), (2) $d_{ZX}$ (green), (3) $d_{YZ}$ (cyan), (4) $d_{X^2-Y^2}$ (blue) and (5) $d_{XY}$ (pink).
\label{fig_fs}}
\end{center}
\end{figure}

For our discussion, we analyze the following model consists of the Fe $3d$-orbitals,
\begin{equation}
H=H_0+H_\mathrm{int}. \label{eq_H}
\end{equation}
 Here, the kinetic part $\hat{H}_0$ is determined so as to reproduce the first-principles band structure for LaFeAsO\cite{kuroki} and its FSs, the band structure and the orbital weight on the FSs are shown in Fig. \ref{fig_fs}.
In present paper, we set $d_{3Z^2-R^2}, d_{ZX}, d_{YZ}, d_{X^2-Y^2},$ and $d_{XY}$ orbitals as 1,2,3,4 and 5 where $x, y$ axes ($X, Y$ axes) are along the nearest Fe-Fe (Fe-As) direction. The Coulomb interaction part is given as
\begin{eqnarray}
H_{\rm int}&=&\frac{1}{2}U\sum_{i}\sum_{\ell}\sum_{\sigma\neq\bar{\sigma}}
d^{\dag}_{i\ell\sigma}d^{\dag}_{i\ell\bar{\sigma}}
d_{i\ell\bar{\sigma}}d_{i\ell\sigma} \nonumber \\
&+&\frac{1}{2}U'\sum_{i}\sum_{\ell\neq\bar{\ell}}\sum_{\sigma,\sigma'}
d^{\dag}_{i\ell\sigma}d^{\dag}_{i\bar{\ell}\sigma'}
d_{i\bar{\ell}\sigma'}d_{i\ell\sigma} \nonumber \\
&+&\frac{1}{2}J\sum_{i}\sum_{\ell\neq\bar{\ell}}\sum_{\sigma,\sigma'}
d^{\dag}_{i\ell\sigma}d^{\dag}_{i\bar{\ell}\sigma'}
d_{i\ell\sigma'}d_{i\bar{\ell}\sigma} \nonumber \\
&+&\frac{1}{2}J'\sum_{i}\sum_{\ell\neq\bar{\ell}}\sum_{\sigma\neq\bar{\sigma}}
d^{\dag}_{i\ell\sigma}d^{\dag}_{i\ell\bar{\sigma}}
d_{i\bar{\ell}\bar{\sigma}}d_{i\bar{\ell}\sigma}, \label{eq_H_int}
\end{eqnarray}
which includes the multi-orbital interaction on a Fe site: the intra- and inter-orbital direct terms $U$ and $U'$, the Hund's rule coupling $J$ and the pair transfer $J'$.

To solve the model, we use the DMFT\cite{georges} which approximates the lattice model by a single-site problem of electrons in an effective medium that may be described by the frequency dependence. In the actual calculations with the DMFT, we solve the effective five-orbital impurity Anderson model, where the Coulomb interaction at the impurity site is given by the same form as $\hat{H}_{\rm int}$ with a site $i$, and the kinetic energy responsible for the bare impurity Green's function ${\cal\hat{G}}$ in the $5\times 5$ matrix representation is determined so as to satisfy the self-consistency condition as possible. We use the exact diagonalization (ED) method for a finite-size cluster as an impurity solver to obtain the local quantities such as the self-energy $\hat{\Sigma}$. To avoid CPU-time consuming calculation, we employ the clusters with the site number $N_s=4$ within a restricted Hilbert space, as used in our previous paper\cite{ishizuka_2}; where we approximate the clusters with those of $d_{3Z^2-R^2}$ and $d_{XY}$ orbital by $N_s=2$ since the two orbitals are far from the Fermi energy in contrast to the another three orbitals. We have confirmed that the results with $N_s=4$ are qualitatively consistent with those with $N_s=2$\cite{ishizuka} and quantitatively improved especially for the intermediate interaction regime. Moreover, the studies by the slave-spin mean field\cite{yi_2,medici,medici_2}, the slave-boson mean field (Gutzwiller)\cite{Hardy} approximations, and also the DMFT with the continuous-time quantum Monte Carlo method (CT-QMC)\cite{aichhorn} give a similar results over our approach. Then, we expect that the present calculation is sufficiently accurate at least up to the intermediate regime.

Within the DMFT, the spin (charge-orbital) susceptibility is given in the $25\times 25$ matrix representation as 
 \begin{eqnarray}
 \hat{\chi}_{s(c)}(q)=
 \left[1 -(+)\hat{\chi}_0(q)\hat{\Gamma}_{s(c)}(i\omega_n)\right]^{-1} 
 \hat{\chi}_0(q),
 \label{eq:chi}
 \end{eqnarray}
with $\hat{\chi}_0(q)=-(T/N)\sum_{k}\hat{G}(k+q)\hat{G}(k)$, where $\hat{G}(k)=[(i\varepsilon_m+\mu)-\hat{H}_0(\bm{k})-\hat{\Sigma}(i\varepsilon_m)]^{-1}$ is the lattice Green's function, $\hat{H}_0(\bm{k})$ is the kinetic part of the Hamiltonian with the wave vector $\bm{k}$, $\hat{\Sigma}(i\varepsilon_m)$ is the lattice self-energy, which coincides with the impurity self-energy obtained in impurity Anderson model, and $k=(\bm{k},i\varepsilon_m)$, $q=(\bm{q},i\omega_n)$. Here, $\varepsilon_m=(2m+1)\pi T$ and $\omega_n=2n\pi T$ are fermionic and bosonic Matsubara frequencies.  In eq. (\ref{eq:chi}), $\hat{\Gamma}_{s(c)}(i\omega_n)$ is the local irreducible spin (charge-orbital) vertex in which only the external frequency ($\omega_n$) dependence is considered as a simplified approximation\cite{ishizuka_2} and is explicitly given by
\begin{eqnarray}
\hat{\Gamma}_{s(c)}(i\omega_n)=-(+)\left[\hat{\chi}_{s(c)}^{-1}(i\omega_n)-\hat{\chi}_0^{-1}(i\omega_n)\right],
\label{eq:gamma}
\end{eqnarray} 
with
$
\hat{\chi}_0(i\omega_n)=-T\sum_{\varepsilon_m}
\hat{G}(i\varepsilon_m+i\omega_n)\hat{G}(i\varepsilon_m), 
$
where $\hat{\chi}_{s(c)}(i\omega_n)$ is the local part of spin (charge-orbital) susceptibility. When the largest eigenvalue $\alpha_s(\bm{q})$ [$\alpha_c(\bm{q})$] of  $(-)\hat{\chi}_0(q)\hat{\Gamma}_{s(c)}(i\omega_n)$ in eq. (\ref{eq:chi}) for a wave vector $\bm{q}$ with $i\omega_n=0$ reaches unity, the instability towards the magnetic (charge-orbital) order with the corresponding $\bm{q}$ takes place. After convergence of the DMFT self-consistent loop, the quantity $\hat{\chi}_{s(c)}(i\omega_n)$ in eq. (\ref{eq:gamma}) is measured by means of continued fraction algorithm \cite{georges}. It includes automatically all vertex corrections respect to the $\hat{\chi}_0(i\omega_n)$, without the need of explicit calculation of the local irreducible vertex function.

The effective pairing interaction for the spin-singlet state, obtained by using the spin (charge-orbital) susceptibility in eq. (\ref{eq:chi}) and the spin (charge-orbital) vertex in eq. (\ref{eq:gamma}), is given as
 \begin{align}
 \hat{V}(q)
 &=\frac{3}{2}\hat{\Gamma}_{s}(i\omega_n)\hat{\chi}_{s}(q)\hat{\Gamma}_{s}(i\omega_n)
 -\frac{1}{2}\hat{\Gamma}_{c}(i\omega_n)\hat{\chi}_{c}(q)\hat{\Gamma}_{c}(i\omega_n) \nonumber \\
 &+\frac{1}{2}\left(\hat{\Gamma}_{s}^{(0)}+\hat{\Gamma}_{c}^{(0)}\right),
 \label{eq:pair}
 \end{align}
 with the bare spin (charge-orbital) vertex: 
 $[\Gamma_{s(c)}^{(0)}]_{\ell\ell\ell\ell}=U(U)$, 
 $[\Gamma_{s(c)}^{(0)}]_{\ell\ell'\ell\ell'}=U'(-U'+2J)$, 
 $[\Gamma_{s(c)}^{(0)}]_{\ell\ell\ell'\ell'}=J(2U'-J)$ and 
 $[\Gamma_{s(c)}^{(0)}]_{\ell\ell'\ell'\ell}=J'(J')$, 
 where $\ell'\neq \ell$ and the other matrix elements are 0. Substituting the effective pairing interaction in eq. (\ref{eq:pair}) into the linearized Eliashberg equation: 
 \begin{align}
 \lambda \Delta_{ll'}(k)&=-\frac{T}{N}\sum_{k'}
 \sum_{l_1l_2l_3l_4}V_{ll_1,l_2l'}(k-k') \nonumber \\
 &\times G_{l_3l_1}(-k')\Delta_{l_3l_4}(k') G_{l_4l_2}(k'),
 \label{gapeq}
 \end{align}
 we obtain the gap function $\Delta_{ll'}(k)$ with the eigenvalue $\lambda$ which becomes unity at the superconducting transition temperature $T=T_c$. To solve eq. (\ref{gapeq}), we neglect the frequency dependence of the vertex: ${\hat \Gamma}_{s(c)}(i\omega_n)\approx{\hat \Gamma}_{s(c)}(i\omega_n=0)$ as a simple approximation but the effect of the frequency dependence will be discussed later.

All calculations are performed for the electron number $n=6.0$ corresponding to the non-doped case at $T=0.02{\rm eV}$ except for the ED calculation in the impurity Anderson model where we calculate the self-energy at $T=0$ as the explicit $T$-dependence is expected to be small at low temperature $T=0.02$eV in the intermediate correlation regime with $Z\simg0.5$. We use $32\times 32$ ${\bm k}$-point meshes and 1024 Matsubara frequencies in the numerical calculations with the fast Fourier transformation. Here and hereafter, we measure the energy in units of eV.

\section{Numerical Results}
In the previous paper\cite{ishizuka_2}, It was shown that, for $U>U'$, the $s_{\pm}$-pairing is realized by the magnetic fluctuation near the AFM order, while, for $U<U'$, the $s_{++}$-pairing is realized by the orbital fluctuation near the FO order within the DMFT+Eliashberg equation. In the present paper, we focus on the typical parameter $U\sim U'$ by putting $U=U'-0.2$ and $J=J'=0.15$ corresponding to intermediate region of $U>U'$ and $U<U'$, where the magnetic and the orbital fluctuations coexist, although itinerant metal should satisfy the relations $U=U'+2J$.

\subsection{Renormalization factor}
\begin{figure}
\begin{center}
\includegraphics[width=74mm]{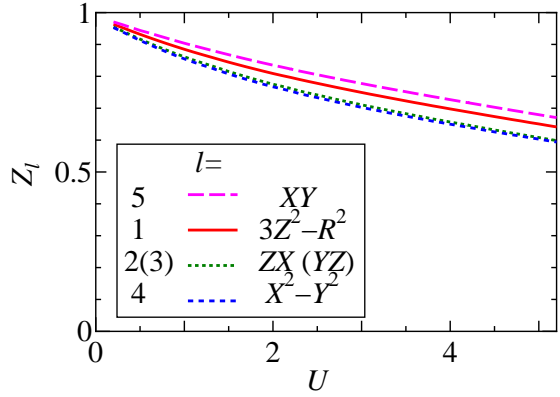}
\caption{(Color online)
The renormalization factor $Z_{\ell}$ with $\ell=d_{3Z^2-R^2}, d_{ZX}, d_{YZ}, d_{X^2-Y^2},$ and $d_{XY}$ as functions of $U$ with $U=U'-0.2$, $J=0.15$ and $J=J'$ for $n=6.0$ and $T=0.02$. The orbital numbers are the same as Fig. \ref{fig_fs}.
\label{fig_z}}
\end{center}
\end{figure}

First, we discuss the self-energy correction. Figure \ref{fig_z} shows the renormalization factor $Z_{\ell}=\left[1-\frac{d\Sigma_{\ell}(\varepsilon)}{d(\varepsilon)}\bigl.\bigr|_{\varepsilon\rightarrow0}\right]^{-1}$, as functions of $U$. When $U$ increases, all of $Z_{\ell}$ gradually decrease with the weak orbital dependence. The orbital dependence of the renormalization factor largely depends on the Hund's coupling $J$ and crystal field splitting of the five $d$ orbitals, as previously discussed by several authors\cite{koga,ferrero,medici_2}. The Hund's coupling stabilizes orbital selective Mott phase (OSMP) since $J$ enhances (suppresses) magnetic (orbital) fluctuations\cite{koga}. More generally, the imbalance between the intra- and the inter-orbital Coulomb interaction is critical for OSMP. Indeed, the small $Z_{\ell}$ of the $X^2-Y^2$ orbital is found for the both sides of $U>U'$ and $U<U'$ in our previous paper\cite{ishizuka_2}. On the other hands, when the orbital and magnetic fluctuation are competing, the metallic state with almost orbital-independent $Z_\ell$ is stabilized\cite{koga}. This is the reason why $Z_{\ell}$ shows the small orbital dependence as shown in Fig. \ref{fig_z}.

\subsection{Vertex function}
Next, we address the irreducible two-particle vertex function. We consider the independent eight of ${\hat \Gamma}_\xi$, i.e. the four different orbital combinations [$(\ell\ell\ell\ell),(\ell\ell'\ell\ell'),(\ell\ell\ell'\ell'),(\ell\ell'\ell'\ell)$] for each of the two channels [$\xi=(s,c)$] and the other matrix elements are neglected.

\begin{figure}
\begin{center}
\includegraphics[width=74mm]{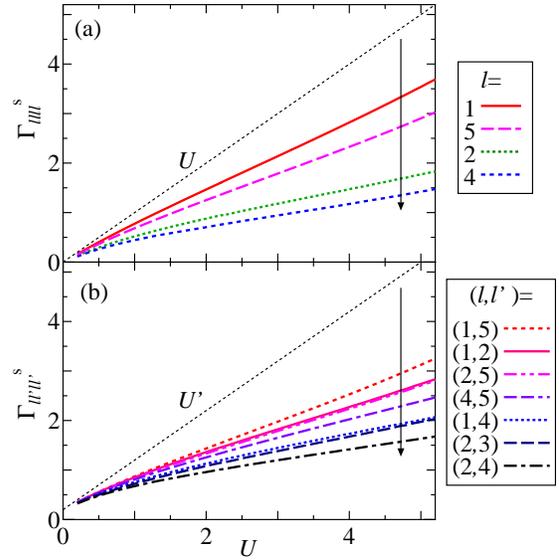}
\caption{(Color online)
The spin vertex ${\hat \Gamma}^s$ for the orbital-diagonal components (a) and the orbital-off-diagonal components (b) with the lowest Matsubara frequency $i\omega_n=0$ as functions of $U$. The bare vertex are also plotted by thin-dot lines. The orbital numbers are the same as Fig. \ref{fig_fs}.
\label{fig_gam1}}
\end{center}
\end{figure}

Figures \ref{fig_gam1}(a) and (b) show the spin vertex $\hat{\Gamma}_{s}(i\omega_n=0)$ with several orbitals, where the zeroth-order contribution of the vertex represents the bare Coulomb repulsion (thin-dot lines). We find that $\hat{\Gamma}_{s}$ is strongly renormalized with the significant orbital dependence as $U$ increases. The orbital-diagonal components of the spin vertex is renormalized with $d_{X^2-Y^2}, d_{ZX}, d_{YZ}, d_{XY}$ and $d_{3Z^2-R^2}$ orbital in ascending order to eliminate the magnetic instability. The $X^2-Y^2$ orbital has the weakest $\hat{\Gamma}_{s}$ because of the large weights of its single-particle density of states, while the spin vertex related to the $XY$ and/or the $3Z^2-R^2$ orbitals which have the small weights on FSs shows the weak renormalization relative to the other orbitals [see also Fig. \ref{fig_fs}(c)-(f)]. The orbital dependence of the vertex (and also the renormalization factor) may be described by perturbation theory in the weak coupling regime.

\begin{figure}
\begin{center}
\includegraphics[width=74mm]{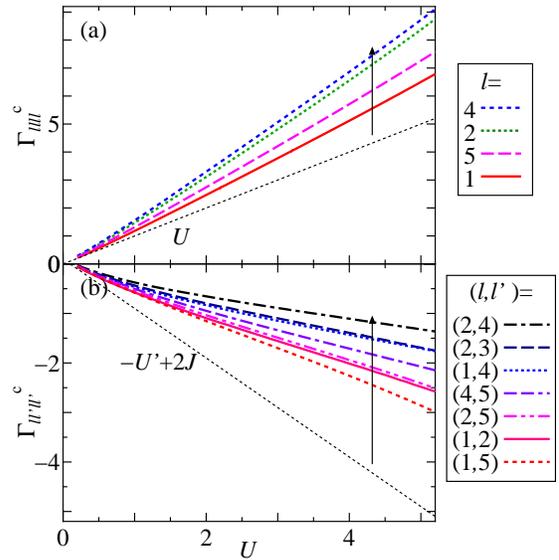}
\caption{(Color online)
The charge-orbital vertex ${\hat \Gamma}^c$ for the orbital-off-diagonal components of (a) and the orbital-off-diagonal components (b) with the lowest Matsubara frequency $i\omega_n=0$ as functions of $U$. The bare vertex are also plotted by thin-dot lines. The orbital numbers are the same as Fig. \ref{fig_fs}.
\label{fig_gam2}}
\end{center}
\end{figure}

In the direct contrast to the spin vertex, the charge vertex is enhanced by the correlation effect [Fig. \ref{fig_gam2}(a)] which makes the charge fluctuation smaller. The enhancement is qualitatively consistent with the single-orbital DMFT+ED study\cite{rohringer}. On the other hands, the orbital-off-diagonal components of the charge vertex is strongly renormalized similar to the case of the spin vertex [Fig. \ref{fig_gam2}(b)], since the orbital and the magnetic fluctuations are mutually suppressed by correlation effects.

The deviation between the orbital-diagonal components of the spin and the charge vertex shows $\Gamma_s<\Gamma^{(0)}<\Gamma_c$ which is qualitatively consistent with the self-consistent fluctuation theory\cite{kusunose_2}. It indicates that applying the weak coupling theory, such as the HF-RPA theory, the same parameter of the Coulomb repulsion in the spin and the charge vertex should not be used even in the single-orbital model. The irreducible two-particle vertex function may be used to build up the weak coupling calculation for parametrizing the vertex functions.

In Ref. \citenum{kanamori}, effective Coulomb repulsion which could achieves an emergent condition of the magnetic instabilities is approximately estimated by a screening effect of particle-particle scattering: $U^{\rm eff}=U[1+U\psi_0(0)]^{-1}$, where $\psi_0(q)=T/N\sum_kG(k)G(q-k)$. This relation is justified in the dilute electron gas. It should be noticed that the renormalization of the vertex functions, in principle, partly includes the particle-particle multiple scattering discussed in Ref. \citenum{kanamori}.

\subsection{Spin and charge-orbital Stoner factors}
\begin{figure}
\begin{center}
\includegraphics[width=74mm]{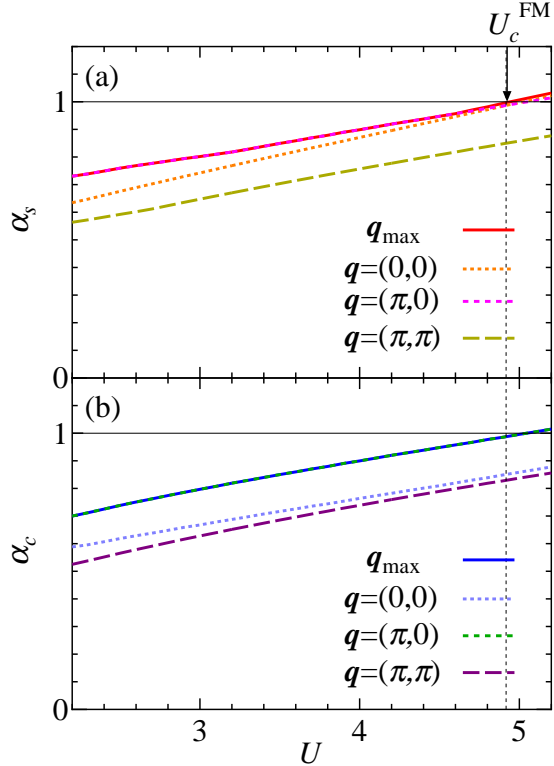}
\caption{(Color online)
(a) and (b) the largest eigenvalues $\alpha_s$ and $\alpha_c$ for several ${\bm q}$ and $\lambda$ which reach unity towards the magnetic, charge-orbital and superconducting instabilities, respectively, as functions of $U$ with $U=U'-0.2$, $J=0.15$ and $J=J'$ for $n=6.0$ and $T=0.02$.
\label{fig_stoner}}
\end{center}
\end{figure}

$U$-dependence of the largest eigenvalues $\alpha_{s}(\bm{q})$ and $\alpha_{c}(\bm{q})$ is calculated and plotted for several wave vectors ${\bm q}$ in Fig. \ref{fig_stoner}(a)-(b) where $\alpha_{s(c)}(\bm{q})$ shows the maximum at ${\bm q}={\bm q}_{\rm max}$. Within the HF-RPA theory, the critical interaction of magnetic instability is given by $U_c\sim0.8$, while the instability in the present study are largely suppressed as $U_c\sim 5$ because of the self-energy and the vertex correction in the strong correlation regime, where the magnetic and the orbital ordered states are competing. The AFM and the AFO eigenvalues of $\alpha_{s(c)}$ are dominant in wide parameter region. However, when $U$ increase, the FM eigenvalue of $\alpha_{s}$ becomes competitive from the AFM and the AFO one, and finally becomes unity. The FM fluctuation originates from the two-effects: (i) the ${\bm q}\sim\bm0$ nesting between the inter-hole FS1 with the large $ZX/YZ$ orbital weights [see Fig.\ref{fig_fs}(c)] and the outer-hole FS2 with the $XY$ orbital weight [see Fig.\ref{fig_fs}(d)], (ii) the weak renormalization of the spin vertex with $d_{ZX}$-$d_{XY}$ orbital components as shown in Fig. \ref{fig_gam1}(b).

To clarify the effect of the vertex correction on the magnetic and the orbital fluctuations, we compare the specific case as follows. (i) (second row of Table \ref{table_compare}) We replace the vertex functions with the bare vertex functions which is corresponding to a RPA-type calculation except for the self-energy correction $[\Gamma_{s(c)}]_{\ell_1\ell_2\ell_3\ell_4}\approx[\Gamma^{(0)}_{s(c)}]_{\ell_1\ell_2\ell_3\ell_4}$, the AFM fluctuation is largely enhanced as compared with the other three fluctuations as the strongest renormalization of the orbital-diagonal components of the spin vertex, which enhances the AFM fluctuation, is neglected. (ii) (third row of Table \ref{table_compare}) Then, if we replace the vertex function with the orbital-{\it independent} vertex functions, i.e. orbital-averaged in each eight-channels $[\Gamma_{s(c)}]_{\ell_1\ell_2\ell_3\ell_4}\approx[\bar{\Gamma}_{s(c)}]_{\ell_1\ell_2\ell_3\ell_4}$, the AFO fluctuation become comparable with the AFM fluctuation. The orbital-diagonal components of the spin and the charge vertex fill a gap between the magnetic and the charge-orbital fluctuation, while the orbital-off-diagonal one do not because of the following reason. The enhancement of the orbital-diagonal components of the charge vertex is moderate as compared with the renormalization of the spin vertex $[\bar{\Gamma}_c]_{\ell\ell\ell\ell}/U\sim1.5$ and $[\bar{\Gamma}_s]_{\ell\ell\ell\ell}/U\sim0.5$. In contrast to the orbital-diagonal components, the orbital-off-diagonal components of the charge vertex is renormalized similar to the spin vertex $[\bar{\Gamma}_c]_{\ell\ell'\ell\ell'}/(-U'+2J)\sim0.46$ and $[\bar{\Gamma}_s]_{\ell\ell'\ell\ell'}/U'\sim0.51$. Thus the orbital-{\it independent} vertex renormalization lifts a relative level of the magnetic and the charge-orbital fluctuations due to the orbital-diagonal components of vertex functions. (iii) (fourth row of Table \ref{table_compare}) Simultaneously, the effect of the orbital-dependent vertex corrections indicates that FM fluctuation is stabilized because of the presence of the small renormalization of the spin vertex in $XY$ orbital as discussed above. The result shows that the orbital degrees of freedom makes an important rule to stabilize the FM fluctuation. Then, we conclude that the major factor of the FM fluctuation is the orbital dependence of the vertex correction.

\begin{table}
\begin{center}
\begin{tabular}{ccccc} \hline\hline
calculation conditions            & $\alpha_s^{\rm AFM}$ & $\alpha_s^{\rm FM}$ & $\alpha_c^{\rm AFO}$ & $\alpha_c^{\rm FO}$ \\ \hline
$[\Gamma_{s(c)}]_{\ell_1\ell_2\ell_3\ell_4}\approx[\Gamma^{(0)}_{s(c)}]_{\ell_1\ell_2\ell_3\ell_4}$ & 1.000  & 0.498  & 0.727  & 0.586 \\
$[\Gamma_{s(c)}]_{\ell_1\ell_2\ell_3\ell_4}\approx[\bar{\Gamma}_{s(c)}]_{\ell_1\ell_2\ell_3\ell_4}$ & 1.000  & 0.766  & 0.966  & 0.785 \\
$[\Gamma_{s(c)}]_{\ell_1\ell_2\ell_3\ell_4}$                                                        & 1.000  & 1.003  & 1.001  & 0.863 \\ \hline\hline
\end{tabular}
\caption{Spin and charge-orbital Stoner factors for ${\bm q}=(\pi,0)$, ${\bm q}=(0,0)$. The calculation conditions with (second row) the RPA-type vertices (neglect the vertex correction) for $U=0.76$, (third row) the orbital-independent vertex functions (averaged in orbitals) for $U=2.66$  and (fourth row) the vertex correction (normal DMFT) for $U=5.05$.
\label{table_compare}}
\end{center}
\end{table}

\subsection{Susceptibility and effective pairing interaction}
\begin{figure}
 \begin{center}
 \includegraphics[width=74mm]{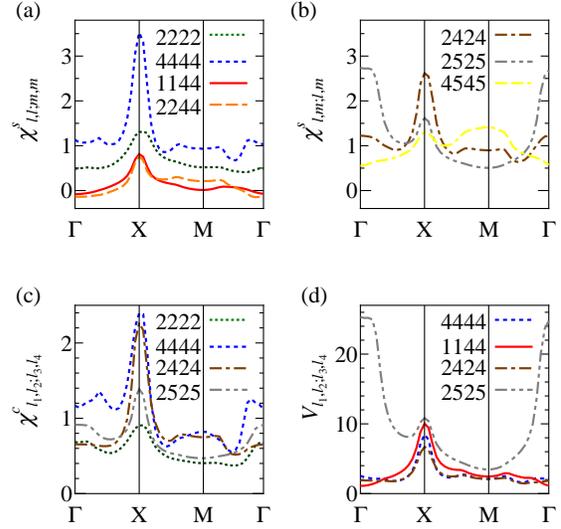}
 \caption{(Color online)
 The spin susceptibility $\chi^{s}_{\ell,\ell;m,m}$ (a), $\chi^{s}_{\ell,m;\ell,m}$ (b), the orbital susceptibility ${\hat \chi^{c}}$(c) and the pairing interaction ${\hat V}$ for several orbital components in ${\bm q}$ space with the lowest Matsubara frequency $i\omega_n=0$ for $U=4.5$, $U'=4.7$ and $J=J'=0.15$, where $\alpha_s=0.958$ and $\alpha_c=0.958$ for ${\bm q}_{\rm max}$. The orbital numbers are the same as Fig. \ref{fig_fs}.
 \label{fig_sus1}}
 \end{center}
 \end{figure}

Figures \ref{fig_sus1} (a) and (b) show the intra- and the inter-orbital components of the spin susceptibility $\chi^{s}_{\ell,\ell;m,m}$ and $\chi^{s}_{\ell,m;\ell,m}$ with the lowest Matsubara frequency $i\omega_n=0$ for $U=4.5$, $U'=4.7$ and $J=J'=0.15$, where the spin Stoner factor is $\alpha_{s}=0.958$. $\chi^{s}_{4,4;4,4}$ around ${\bm q}\sim (\pi,0)$ is enhanced by the effect of the intra-orbital nesting between the hole FS3 and the electron FS, where the $d_{X^2-Y^2}$ component has a large contribution to the density of state in the both FSs as shown in Fig. \ref{fig_fs}(e) and (f). That is, the spin susceptibility $\sum_{\ell,m}\chi^{s}_{\ell,\ell;m,m}$ develop mainly on the $d_{X^2-Y^2}$ orbital. Note that the inter-orbital spin susceptibility $\chi^{s}_{2,4;2,4}$ is also large for ${\bm q}\sim (\pi,0)$ [Fig. \ref{fig_sus1}(b)].

Remarkably, $\chi^{s}_{2,5;2,5}$ [Fig.\ref{fig_sus1}(b)] around the ${\bm q}\sim \bm0$ is enhanced by the inter-orbital nesting between the inter-hole and the outer-hole FSs [Fig.\ref{fig_fs}(c) and (d)]. The enhancement of FM fluctuation is due to the vertex correction on the spin vertex in $d_{ZX}$-$d_{XY}$ orbital which is neglected in the RPA [see also Fig.\ref{fig_gam2}(b) $(\ell,\ell')=(2,5)$].

Figure \ref{fig_sus1} (c) shows the intra- and inter-orbital components of the orbital susceptibility $\chi^{c}$. The charge-orbital Stoner factor is $\alpha_{c}=0.958$. Similar to the spin susceptibility, the orbital susceptibility $\chi^{s}$ around ${\bm q}\sim (\pi,0)$ is enhanced by the intra- and inter-orbital nesting effects. In present model, one observes $\chi^{c}_{4,4;4,4} \approx \chi^{c}_{2,4;2,4}$.

Figure \ref{fig_sus1} (d) show the obtained pairing interaction $V$. The strong enhancement of $V_{2,5;2,5}$ for ${\bm q}\sim \bm0$ is observed due to the FM fluctuation, whereas the moderate enhancement of $V$ for ${\bm q}\sim (\pi,0)$ is observed due to the AFM fluctuation. This is because the attractive AFO fluctuation cancel out the repulsive AFM one in the pairing channel.

\subsection{Gap function}
\begin{figure}
 \begin{center}
 \includegraphics[width=74mm]{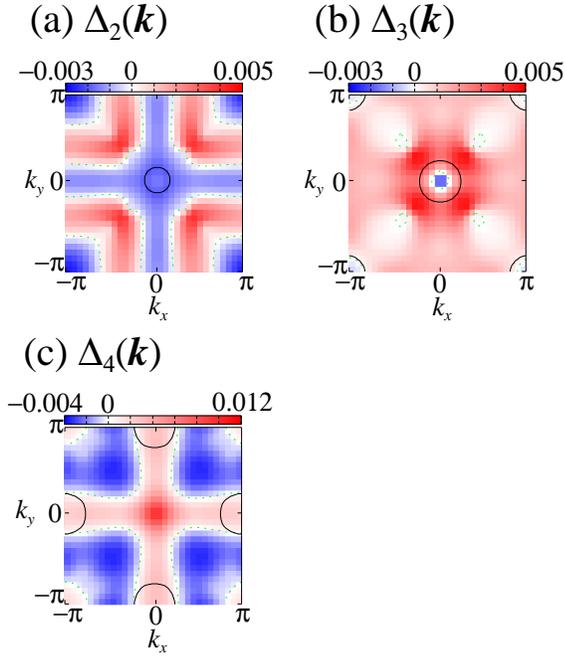}
 \caption{(Color online)
 The band-diagonal components of the gap function $\Delta$ with the lowest Matsubara frequency $i\varepsilon_m=i\pi T$ for band 2 (a) and band 3 (b) (band 4 (c)) with the hole (electron) FSs (solid lines) for $U=4.5$, $U'=4.7$ and $J=J'=0.15$, where the eigenvalue $\lambda=1.07$.
 \label{fig_gap1}}
 \end{center}
 \end{figure}

Next we discuss the superconducting state when the magnetic and the orbital fluctuation coexist. In Figs. \ref{fig_gap1} (a)-(c), we show the gap function $\Delta$ with the lowest Matsubara frequency $i\varepsilon_m=i\pi T$ for band 2-4 where the hole (electron) FSs are also plotted . It is shown that the hole-$s_{\pm}$-wave state mediated by the FM fluctuation which favors the sign change of the gap function between the inner hole-pocket and outer hole-pocket is realized. It is worth to notice that the gap function on $ZX/YZ$ orbitals in orbital representation has opposite sign with $XY$ orbital in the momentum space (not shown). Our obtained superconducting symmetry is the same sign between the hole FS3 and the electron FS and the opposite sign between the inner-hole FS1 and the outer-hole FS2, and is summarized as $(\Delta_{h1},\Delta_{h2},\Delta_{h3},\Delta_{e})=(-,+,+,+)$. We argue that the hole-$s_{\pm}$-wave state shown in Figs. \ref{fig_gap1} (a)-(c) is different from the previously proposed states\cite{ahn,saito,yin}, where the superconducting gap is given by $(\Delta_{h1},\Delta_{h2},\Delta_{h3},\Delta_{e})=(-,+,-,+),(+,+,-,+),(+,+,-,-)$. However, the absolute value and the anisotropy of the gap function is similar to Ref. \citenum{saito}.

Since the most iron-based superconductor is considered to be the $s_{\pm}$-wave pairing or the $s_{++}$-wave pairing, the hole-$s_{\pm}$-wave pairing is a exotic superconducting state. The hole-$s_{\pm}$-wave state (and orbital antiphase $s_{\pm}$-wave state) have been proposed in the literature\cite{ahn,saito,yin}, but the correlation-induced hole-$s_{\pm}$-wave state is more pronounced in the present case than in the above two cases. As is evident in the angle-resolved photoemission spectroscopy\cite{borisenko}, a nodeless gap structure of the hole-$s_{\pm}$-wave state is consistent with LiFeAs. Although the quasiparticle interference experiments\cite{hanke,allan,hess} can, in principle, determine relative signs of the gaps on various FSs, the hole-$s_{\pm}$-wave state can not be distinguished from other $s$-wave state experimentally.

\begin{figure}
 \begin{center}
 \includegraphics[width=74mm]{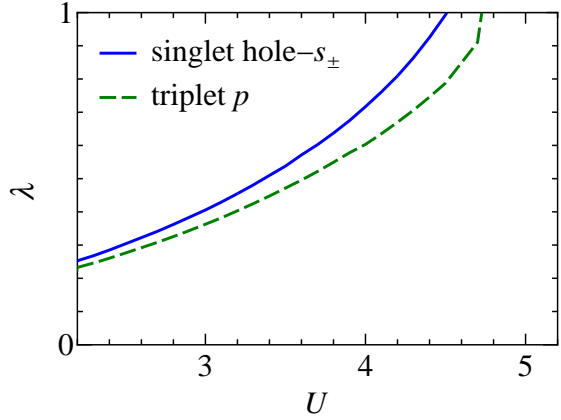}
 \caption{(Color online)
 The eigenvalue $\lambda$ of Eq. \ref{gapeq} as functions of $U$ with $U=U'-0.2$, $J=0.15$ and $J=J'$ for $n=6.0$ and $T=0.02$. The hole-$s_{\pm}$-wave state is realized for a wide region.
 \label{fig_lambda}}
 \end{center}
 \end{figure}

Ref. \citenum{brydon} discussed the spin-triplet chiral $p$ state in three band model for LiFeAs due to the weak nesting effects. Thus we discuss the possibility of a spin-triplet superconducting state. Within the DMFT+Eliashberg equation, the effective pairing interaction for the spin-triplet state is given as $\hat{V}(q)=-\frac{1}{2}\hat{\Gamma}_{s}\hat{\chi}_{s}(q)\hat{\Gamma}_{s}-\frac{1}{2}\hat{\Gamma}_{c}\hat{\chi}_{c}(q)\hat{\Gamma}_{c}+\frac{1}{2}(\hat{\Gamma}_{s}^{(0)}+\hat{\Gamma}_{c}^{(0)})$. Here, the electron-electron Coulomb interaction is symmetric under rotations in spin space. Therefore, pairing interaction $\hat{V}(q)$ induced only by the Coulomb repulsion does not bring about any anisotropy in spin space, and never lifts the degeneracy of the $d$-vector states in principle. Then we calculate only $p_x$-type solution of the gap function. The spin-triplet superconducting eigenvalue is not larger than the singlet one, but appreciably enhanced by the Coulomb interaction (see Fig.\ref{fig_lambda}).

\section{Summary and discussion}
In summary, we have investigated superconductivity in five-orbital Hubbard model for iron pnictides for the case that the intra-orbital Coulomb interaction $U$ is set to be a little smaller than the inter-orbital one $U'$ with a fixed Hund's coupling $J$, by using the DMFT+ED method. We have found that when $U$ increases, the FM fluctuation of inter-orbital component is slowly enhanced because the vertex correction leads to suppression of AFM and AFO order. It seems to be consistent with LiFeAs\cite{wright_2} where both FM and AFM fluctuations are observed, although the FM fluctuation is observed only in high temperature. To determine the superconducting pairing symmetry, we calculate the pairing interaction mediated by the spin-charge-orbital fluctuations obtained from the susceptibilities and the two-particle vertex function and is substituted it into the linearized Eliashberg equation where the single-particle Green's function is renormalized due to the local self-energy correction within the DMFT. Remarkably, the hole-$s_{\pm}$-wave state mediated by the FM fluctuation accompanied by the AFM fluctuation and the AFO fluctuation is realized in this case.

The HF-RPA theory, in which the self-energy and the vertex correction are neglected, overestimates the magnetic (charge-orbital) fluctuations and the ordered states. To avoid such a overestimation, simply reduced vertex function or the simply renormalized bandstructure are often applied in the weak coupling theory instead of taking into account of the self-energy and the vertex correction. However, the effect of the orbital-independent renormalization of the vertex function fills a gap between the magnetic and the orbital fluctuation, meaning that the inconsistency of the renormalization effect between the spin and the charge vertex as previously discussed in single-orbital Hubbard model\cite{kusunose_2}. Thus one should employs $\Gamma_s<\Gamma^{(0)}<\Gamma_c$ for the orbital-diagonal components as a consequence of the intra-band screening effect in the low-energy downfolding model. In addition, we argue that the orbital-dependent renormalized vertex induces the enhancement of the ferromagnetic (ferro-orbital) fluctuation.

At the large $U$ region, we observed the FM fluctuation with the competition of the AFM and the AFO fluctuation. The existence of the FM fluctuation agrees with other numerical calculations\cite{brydon,platt} for LiFeAs. The reason favoring the ferromagnetism is attributed to, as mentioned above, the weak renormalization of the spin vertex with $d_{ZX}$-$d_{XY}$ orbital components and the ${\bm q}\sim\bm0$ nesting between the inter-hole FS1 with the large $ZX/YZ$ orbital weights and the outer-hole FS2 with the $XY$ orbital weight despite of the existence of the good nesting between the hole and the electron FSs. In the condition of the rotational symmetry $U=U'+2J$, we did not find a signature of the FM fluctuation and the strong vertex correction since the AFO fluctuation is immediately suppressed for $U>U'$. However, if we consider the effect of the coupling $g$ between the electron and the $E_g$ phonon as discussed in Refs. \citenum{kontani} and \citenum{yanagi_1}, the AFO fluctuation can be expanded over the realistic parameter region with $U>U'$. Indeed, a kink structure of the single-particle dispersion around the $\Gamma$ point is observed experimentally in LiFeAs which is due to strong electron-phonon coupling \cite{borisenko}. Thus the strong vertex correction may be realized by taking into account of the realistic electron-phonon coupling.

According to the NMR measurements of LiFeAs\cite{li_4,jeglic,baek_2}, the Knight shift decrease below transition temperature indicating that the spin-triplet state is inconsistent with the experiments. In $H=8.5$T, however, the Knight shift does not show any suppression in $H\perp c$, whereas it decrease in $H\parallel c$\cite{baek_2}. In addition, recent results of detailed field dependence of onset temperature derived from magnetic torque measurement indicate that the unusual spontaneous magnetization in high magnetic field is attributable to the chiral $p$ state of the spin-triplet superconductivity\cite{li_3}. Thus, we expect that the spin-triplet state due to ferromagnetic fluctuation is realized in high magnetic field in LiFeAs.

 In Ref. \citenum{held}, itinerant ferromagnetism has been investigated in order to clarify the electron correlation in the two-orbital degenerated Hubbard model on the basis of the DMFT. The obtained phase diagrams shows that the FM order is stabilized between quarter filling (orbital order) and half filling (AFM order). The important point is that the FM ordered state is stabilized by the Hund's coupling and the band degeneracy. 

Finally, we comment on the dynamical screening of the irreducible vertex functions affecting on the superconducting instabilities. To solve the linearized Eliashberg equation, we replace the frequency-dependent vertex with the constant for simply approximation. However, the frequency dependence in the irreducible vertices is significantly important in strongly correlated electron systems. The dynamical screening of the Coulomb repulsion\cite{takada} influences the retardation effect of the effective pairing interaction which strengthens the pairing magnitude in the high frequency region. In fact, self-consistent renormalization study\cite{nakamura_2} shows that the relative spread of the frequency space of the magnetic fluctuation increases the superconducting transition temperature. Indeed, we have also discussed the approximate effect of the dynamical screening of the irreducible vertex functions on the present model for the iron-based superconductors and have found that the screening effect enhances the magnetic fluctuation-mediated $s$-wave superconductivity. Therefore, we need further investigation of the dynamical screening of the irreducible vertex functions with including the more realistic effects. Generally, the higher order terms may play an important role for the superconducting instability, since it is considered that most of unconventional superconductors are in the intermediate coupling region. For instance, vertex correction due to Aslamazov-Larkin terms which is not included in the RPA plays an important role to stabilize the $s_{++}$-wave state in iron-pnictides\cite{onari}. Therefore, it is an important issue to investigate the role of higher order corrections.

\section*{Acknowledgments}
This work was partially supported by a Grant-in-Aid for Scientific Research from the Ministry of Education, Culture, Sports, Science and Technology and also by Yukawa memorial foundation (Mochizuki fund). 

\bibliographystyle{jpsj.bst}
\bibliography{jpsj2.bib}
\end{document}